\documentclass[a4paper,prl,english,twocolumn,floatfix,amsfonts,amssymb,superscriptaddress,groupedaddress]{revtex4-1}
\usepackage{graphicx}
\usepackage{epstopdf} %converting to PDF
\usepackage{dcolumn}
\usepackage{bm}
\usepackage{bbm,bbold}
\usepackage{color}
\usepackage{xcolor, soul}
\sethlcolor{green}
\usepackage{amsfonts}
\usepackage{amsmath}
\usepackage{mdframed}
\usepackage[normalem]{ulem}
\usepackage{mathrsfs}  

\usepackage[percent]{overpic}
\usepackage[colorlinks=true,citecolor=blue]{hyperref}
\hypersetup{colorlinks=true,citecolor=blue,linkcolor=blue,urlcolor=blue}
\DeclareRobustCommand{\rchi}{{\mathpalette\irchi\relax}}
\newcommand{\irchi}[2]{\raisebox{\depth}{$#1\chi$}} % inner command, used by \rchi

\DeclareMathAlphabet\mathbfcal{OMS}{cmsy}{b}{n}

\begin{document}
\title{Giant Shear Displacement by Light-Induced Raman Force in Bilayer Graphene}
\author{Habib Rostami}
\email{habib.rostami@su.se}

\affiliation{Nordita, KTH Royal Institute of Technology and Stockholm University, Hannes Alfvéns väg 12, 10691 Stockholm, Sweden}
\date{\today}

\begin{abstract}
Coherent excitation of shear phonons in van der Waals layered materials is a non-destructive mechanism to fine-tune the electronic state of the system. We develop a diagrammatic theory for the displacive Raman force and apply it to the shear phonon's dynamics. We obtain a rectified Raman force density in bilayer graphene of the order of ${\cal F}\sim 10{\rm nN/nm^2}$ leading to a giant shear displacement $Q_0 \sim 50$pm for an intense infrared laser. We discuss both circular and linear displacive Raman forces. 
We show that the laser frequency and polarization  can effectively tune $Q_0$ in different electronic doping, temperature, and scattering rates. We reveal that the finite $Q_0$ induces a Dirac crossing pair in the low-energy dispersion that photoemission spectroscopy can probe. Our finding provides a systematic pathway to simulate and analyze the coherent manipulation of staking order in the heterostructures of layered materials by laser irradiation. 
\end{abstract}

\maketitle

\begin{figure}[t]
\centering
\includegraphics[width=6cm]{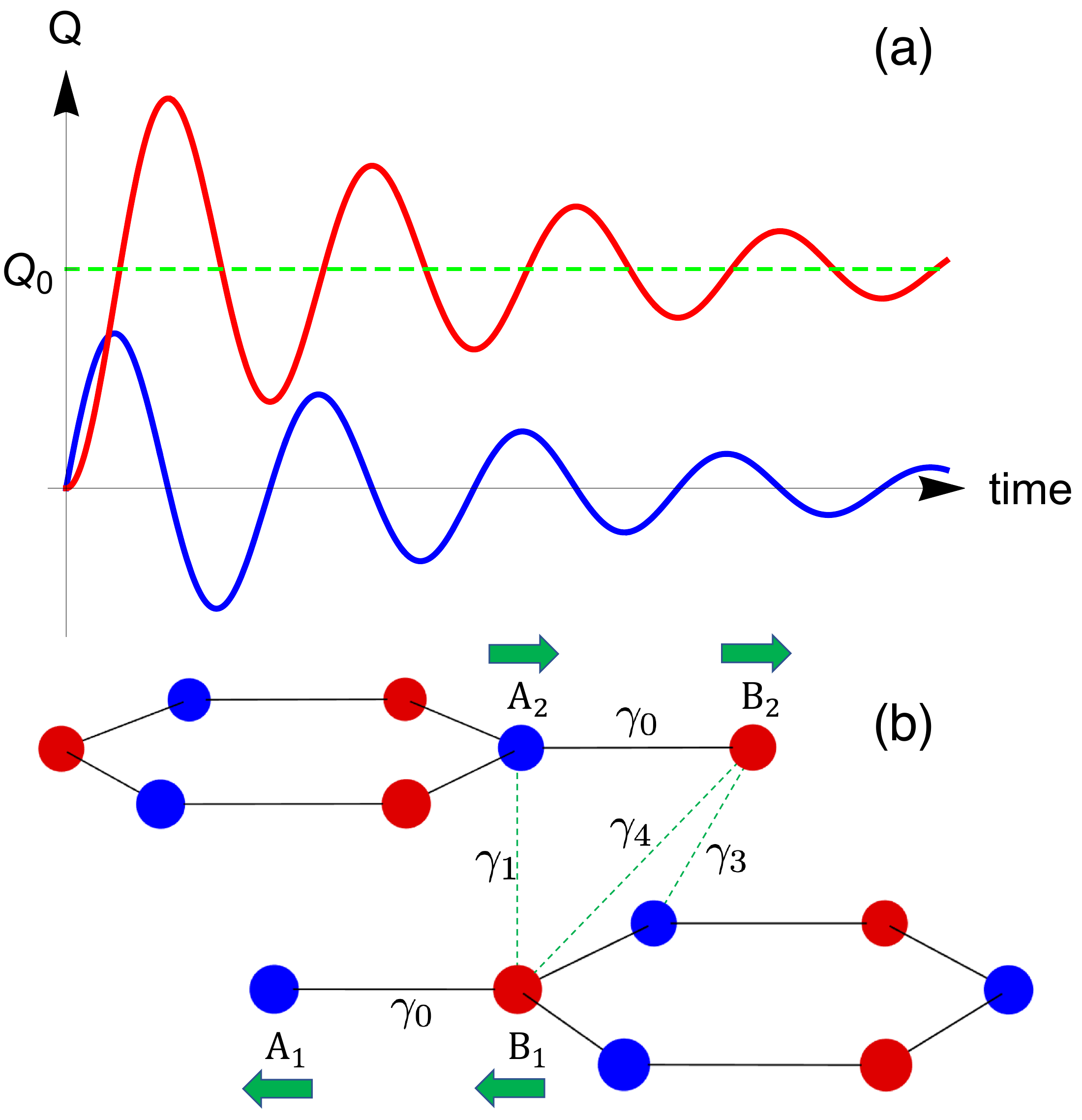}
\caption{{\bf Schematic of coherent shear phonon in bilayer graphene.} (a) Impulsive (blue curve) and displacive (red curve) coherent phonon. (b) Shear phonon mode in bilayer graphene. We illustrate fundamental hopping mechanism in bilayer graphene denoted by $\gamma_i$s. The intralayer Carbon-Carbon bond length is $a_0\approx 0.14$ {\rm nm}, the interlayer distance is $c\approx 0.34 {\rm nm}$ and the bond length for $\gamma_3$ and $\gamma_4$ hoppings is given by $b = \sqrt{c^2+a^2_0}\approx 0.38 {\rm nm}$.
Hopping parameters are given as $\gamma_0 \approx 3$eV, $\gamma_1\approx0.4$eV, $\gamma_3\approx 0.3$eV and $\gamma_4\approx0.1$eV. 
}  
\label{fig:schematic}
\end{figure}

{\em Introduction.--} 
In van der Waals (vdW) layered materials, e.g. the family of graphene and transition metal dichalcogenides (TMDs), the stacking order of layers is crucial for the ground state properties. The properties of an AB stack bilayer graphene (A-sublattice on top of B-sublattice-- see Fig.~\ref{fig:schematic}) are different from that of AA-stack one \cite{Lopes_prb_2007,Ho_prb_2006,Tabert_prb_2012,Bistritzer12233,Lopes_prb_2012,Rakhmanov_prl_2012,McCann_rpp_2013, Roldan_prb_2013}. In trilayer graphene, two common structures are with ABA and ABC stacking with and without center of symmetry, respectively \cite{Koshino_prb_2009,Avetisyan_prb_2009,Avetisyan_prb_2010,li2019electronic,Mora_prl_2019, Phong_prb_2021,Qin_prl_2021}.   In twisted 2D materials with an asymmetric layer rotation, e.g., twisted bilayer and trilayer graphene \cite{Bistritzer12233, Po_prx_2018, Zou_prb_2018, Angeli_prb_2018, Koshino_prx_2018, Kang_prx_2018}, the ground state strongly depends on the twist owing to the flat-band formation at magic twist angles \cite{Bistritzer12233}.  The relative lateral layer shift in a small twist-angle incommensurate bilayers, with a large moir\'e lattice constant, can be gauged away by a unitary transformation \cite{Bistritzer12233}. 
However, its impact is significant for a large twist-angle. A rigid relative displacement along armchair direction by one C-C bond length $a_0\sim 0.14$nm or a $60^\circ$ relative layer twist can switch between AA and AB prototypes. 
The relative lateral displacement of top and bottom layers versus the middle one in twisted trilayer graphene can drastically change the density of states and superconducting critical temperature
\cite{Zhu_prl_2020, Lei_prb_2021, Qin_prl_2021}. Although there is a surge of interest in twisted multilayers, the impact of the relative lateral shift in vdW layered materials is less highlighted in the literature. This work aims to fill this gap.

Shear phonons in vdW layered materials, correspond to the lateral asymmetric sliding of atomic layers \cite{Tan_nm_2012,Ferrari_nn_2013, Zhang_prb_2013,Zeng_prb_2012,Michel_prb_2008,Michel_prb_2012,Zhao_nl_2013,Wang_jpcc_2017, Pizzi_acsnano_2021}. The double degenerate Raman-active shear mode in bilayer graphene has a soft frequency $\hbar\Omega_0\approx3.9$meV \cite{Tan_nm_2012} due to the weak vdW interlayer coupling. In trilayer there are two double degenerate modes: Raman-active $\approx3.9$meV and infrared-active $4.7$meV \cite{Tan_nm_2012}. The stimulated Raman effect is an efficient mechanism to excite Raman-active vibrational modes \cite{dresselhaus2007group,yu2010fundamentals}.  
The photo-induced structural transition in layered quantum materials such as multilayer graphene, WTe$_2$, and MoTe$_2$  is rapidly evolving using the ultrafast pump-probe setup and time-resolved second-harmonic-generation spectroscopy \cite{Sie_nature_2019,Zhang_prx_2019,Fukuda_apl_2020,Zhang_lsa_2020,Ji_acsnano_2021}.   Shear phonon dynamics can manipulate the staking order of layers \cite{Zhang_prx_2019,Ji_acsnano_2021} and the electronic topology \cite{Sie_nature_2019}. 
A displacive coherent excitation of the Raman-active shear phonon in MoTe$_2$ cause a first-order phase transition from inversion symmetric 1T$'$ structure to the non-centrosymmetric 1T$_d$ phase \cite{Zhang_prx_2019,Fukuda_apl_2020}. A structural phase switch from an ABA to ABC stacking is experimentally obtained by laser irradiation on trilayer graphene \cite{Zhang_lsa_2020}. 
  
In this paper, we study coherent shear phonon dynamics employing a diagrammatic framework in vdW layered materials. 
Collective macroscopic oscillation of atoms in a crystalline solid, i.e. coherent phonons, facilitates a non-destructive control of physical properties by irradiating ultrashort laser pulses and employing transient optical spectroscopy \cite{born1955dynamical,lanzani2007coherent,Dekorsy_book_2000,Hase_nature_2003,Ishioka_apl_2006,Zeiger_prb_1992,Pfeifer_1992,Kuznetsov_prl_1994,Kuznetsov_prb_1995,Stevens_prb_2002,Garrett_prl_1996,Merlin_ssc_1997}.
We provide a theory for the displacive Raman force and implement it to excite coherent shear phonon in bilayer graphene (BLG). The excitation of coherent shear modes can change the electronic structure and trigger a structural transition to another quasi-equilibrium state. We obtain highly efficient tunability of Raman force by altering electronic doping and temperature as well as laser frequency, polarization , and power.
The harmonic equation of motion of the coherent phonon displacement ${\bf Q}$ is given by  
\begin{align}\label{eq:eom}
\partial^2_t {\bf Q}(t)+\Omega^2_0 {\bf Q}(t) = \mathbfcal{F}(t)/{\rho},
\end{align}
where $\rho$ stands for the mass density of the 2D material and $\mathbfcal{F}(t)$ is the light-induced force density. 

For an ultrashort $\delta$-function pulse, the Raman force density can have impulsive ${\cal F}^{\rm I}(t)\sim {\cal F} \delta(t)$ and displacive ${\cal F}^{\rm D}(t)\sim {\cal F} \Theta(t)$ character where $\Theta(t)=\int^t_{-\infty} dt' \delta(t')$ is the Heaviside step function \cite{Zeiger_prb_1992,Stevens_prb_2002,Garrett_prl_1996,Merlin_ssc_1997}. The impulsive force leads to the coherent vibration of ions around equilibrium positions while under a displacive force ions shift away from the equilibrium positions to a new local equilibrium and then vibrate around the new equilibrium positions (see Fig.~\ref{fig:schematic}a). The real and imaginary parts of the Raman susceptibility contribute to the impulsive and displasive forces, respectively \cite{Stevens_prb_2002,Garrett_prl_1996,Merlin_ssc_1997}. For the laser frequency larger than the optical transition edge, the imaginary part is finite and thus the displacive Raman force is non-zero \cite{Merlin_ssc_1997}. In what follows,  we evaluate the displacive Raman force and demonstrate its relevance for the light-induced shear displacement in BLG.

{\em Model.--} 
The dipole moment of Raman-active phonon $ \mu_a = \alpha_{ab} E_b$ is linearly proportional to the light electric field $E_b$ where the polarisability tensor $\alpha_{ab}$ depends on the phonon displacement vector ${\bf Q}$. The electromagnetic potential energy then follows $U=-\mu_aE_a= - \alpha_{ab} E_a E_b$. The corresponding Raman force thus reads $F_c= -\partial U/\partial Q_c= \rchi^{\rm R}_{abc} E_aE_b$ with the Raman susceptibility $\rchi^{\rm R}_{abc}=\partial\alpha_{ab}/\partial Q_c|_{Q\to0}$ \cite{dresselhaus2007group,yu2010fundamentals}. We adopt Einstein convention for summation on repeated indices. Although the lowest-order Raman effect is a second-order nonlinear optical process, it does not require an inversion symmetry breaking. This is because the Raman-active mode in a centrosymmetric system is even under parity \cite{yu2010fundamentals}.

The displacive (rectified) force will displace ions to a new equilibrium position   
${\bf Q}_{0} \approx \mathbfcal{F}^{\rm D} /(\rho\Omega^2_0)$.
Then ions vibrate around the new equilibrium with the phonon frequency $\Omega_0$, see Fig.~\ref{fig:schematic}. Apparently, the rigid displacement is more pronounced for the soft shear phonons with shallow frequency relative to other energy scales such as temperature and electronic chemical potential.
The displacive force is governed by the rectification process ${\cal F}^{\rm D}_a = \int dt {\cal F}_a(t) $  and it is given by 
\begin{equation}
 {\cal F}^{\rm D}_a = - \frac{ \chi^{\rm R}_{abc}(\omega,-\omega) }{ \omega^2} 
 E_b(\omega) E^\ast_c(\omega),
\end{equation}
where $\omega$ is the light frequency and $ \chi^{\rm R}_{abc}$ is the Raman response function that is the correlation function of electron-phonon and light-matter couplings.
 \begin{figure*}
\centering
\includegraphics[width=16cm]{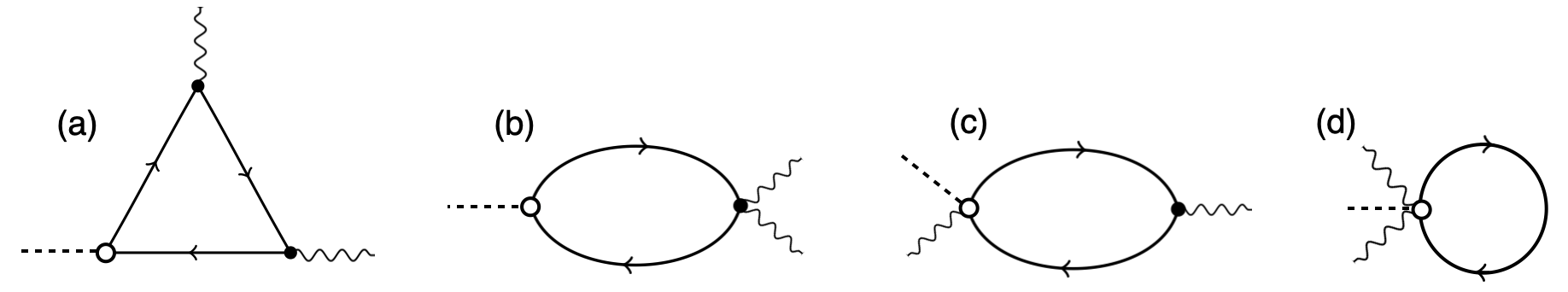}
\caption{{\bf Feynman diagrams for Raman force.} Dashed and wavy lines represent external phonon and photon fields, respectively. The solid lines represent electron propagators.}
\label{fig:diagrams}
\vspace{-3mm}
\end{figure*}

Having the in-plane displacement ${\bf Q}^{(\ell)}({\bf r})$ of two layers $\ell=1,2$, the shear phonon displacement is the asymmetric component: ${\bf Q}= ({\bf Q}^{(1)}-{\bf Q}^{(2)})/{\sqrt{2}}$. The shear displacement vector is even under parity ${\cal P}$ since ${\cal P} \{{\bf Q}^{(1)},{\bf Q}^{(2)}\}{\cal P}^{-1} = -\{{\bf Q}^{(2)},{\bf Q}^{(1)}\}$ that leads to ${\cal P} {\bf Q}{\cal P}^{-1}= {\bf Q}$. Therefore, it is a Raman-active and IR-inactive phonon. The second quantised form of the shear phonon displacement reads 
$\hat Q_{\lambda,{\bf q}} = \sqrt{{\hbar}/{\rho S \Omega_{\lambda,\bf q} }}(\hat b_{\lambda,{\bf q}} + 
\hat b^\dagger_{\lambda,-{\bf q}} )$
where $\lambda$ indicates two Cartesian components $\lambda=x,y$. We recall boson statistics $[\hat b_{\lambda,{\bf q}}, \hat b^\dagger_{\lambda',{\bf q}'}]=\delta_{\lambda\lambda'}\delta_{{\bf q} {\bf q}'}$ and $[\hat b_{\lambda,{\bf q}}, \hat b_{\lambda',{\bf q}'}]=0$. Note that $S$ stands for the area of the 2D material. 
The leading Hamiltonian of finite-$\bf q$ phonon $\hat b_{\lambda,\bf q}$ interacting with electron spinor fields $\hat \Psi_{\bf k}$ is given by  
\begin{align}\label{eq:Ham}
\hat{\cal H} &= \sum_{\bf k} \hat \Psi^\dagger_{\bf k} \hat H_{\bf k} \hat \Psi_{\bf k}
+ \sum_{\lambda,\bf q} \hbar \Omega_{\lambda,\bf q} \hat b^\dagger_{\lambda,\bf q}\hat b_{\lambda,\bf q}
\nonumber\\ &+ 
\sum_{\bf k,q} \sum_{\lambda}\hat\Psi^\dagger_{\bf k+q} \hat g_{\lambda} ({\bf k}, {\bf q}) \hat\Psi_{\bf k} \big (\hat b_{\lambda,\bf q} + \hat b^\dagger_{\lambda,-\bf q} \big), 
\end{align} 
where the electron-phonon coupling is given in terms of the matrix-element $\hat M_{\lambda}$ as $\hat g_{\lambda} ({\bf k}, {\bf q}) =\sqrt{{\hbar}/{2 \rho S \Omega_{\lambda,{\bf q}}}} \hat M_{\lambda}({\bf k},{\bf q})$. 
The hermiticity of the Hamiltonian implies that $\hat M^{\dagger}_\lambda({\bf k}+{\bf q},-{\bf q})  =\hat M_\lambda({\bf k},{\bf q})$. 
Using the Heisenberg time-evolution relation $\langle \partial_t \hat b_{\lambda,{\bf q}} \rangle 
= ({i}/{\hbar}) [\hat {\cal H},\hat b_{\lambda,{\bf q}}] $, we arrive at the classical equation of motion for the coherent phonon $Q_{\lambda,{\bf q}}(t)= \langle \hat Q_{\lambda,{\bf q}} \rangle$. The coherent phonon equation motion follows Eq.~(\ref{eq:eom}) for $\bf q=0$. The Raman force density is given as the expectation value of the electron-phonon coupling
${\cal F}_{\lambda,{\bf q}}(t) = - 
\frac{1}{S}\sum_{\bf k} \langle \hat\Psi^\dagger_{\bf k-q} \hat M_\lambda({\bf k},{-\bf q}) \hat\Psi_{\bf k} \rangle$. 
This force is related to the excitation density, see also Ref.~\cite{Kuznetsov_prb_1995}, where the light-induced electron density generates a force acting on ions. For the normal incidence of light, only $\bf q=0$ phonon is a Raman-active mode.

The light-matter coupling is incorporate by minimal coupling transformation $\hbar {\bf k}\to \hbar {\bf k} + e {\bf A}(t)$ using a homogeneous dynamical vector potential ${\bf A}(t)$. The corresponding electric field reads ${\bf E}(t)= -\partial_t {\bf A}(t)$. The light-matter interaction Hamiltonian consists of two parts: photon-electron terms and photon-electron-phonon terms: 
\begin{align}
&{\cal H}_{lm} = - \sum_{\bf k}  
\hat\Psi^\dagger_{\bf k} 
\Big\{   \hat j_a  A_a(t) 
+ \frac{1}{2} \hat \gamma_{ab}  A_a(t) A_b(t)
\nonumber\\&+   \hat \Theta_{ab}  A_a(t) Q_b(t)  
+\frac{1}{2}   \hat \Delta_{abc}  A_a(t) A_b(t) Q_c(t) \Big\}
\hat\Psi_{\bf k}, 
\end{align}
where $\hat j_a({\bf k})$ is called the paramagnetic current operator and $\hat \gamma_{ab}({\bf k})$ is known as the diamagnetic current operator as well as the Raman vertex in the effective mass approximation \cite{Devereaux_rmp_2007}. The  photon-electron-phonon interaction couplings are parametrised by $\hat \Theta_{ab}({\bf k})$ and $\hat \Delta_{abc}({\bf k})$. The photon-electron-phonon couplings originate from the minimal coupling transformation in the electron-phonon matrix-element $\hat M_\lambda ({\bf k}+e{\bf A}(t)/\hbar,{\bf q})$ and then expanding it up to second order in the light field. We follow the standard many-body perturbation theory and utilize a diagrammatic framework \cite{Rostami_ap_2021,Rostami_npj2dM_2021,Cappelluti_prb_2012b,Cea_prb_2019}. Accordingly, the Raman force response function $\rchi^{\rm R}_{abc}$ consists of four diagrams, as shown in Fig.~\ref{fig:diagrams}. Having defined the key aspects of the model, we calculate the displacive Raman force and the resulting shear displacement in bilayer graphene. 

\begin{figure*}[t]
\centering
\includegraphics[width=16cm]{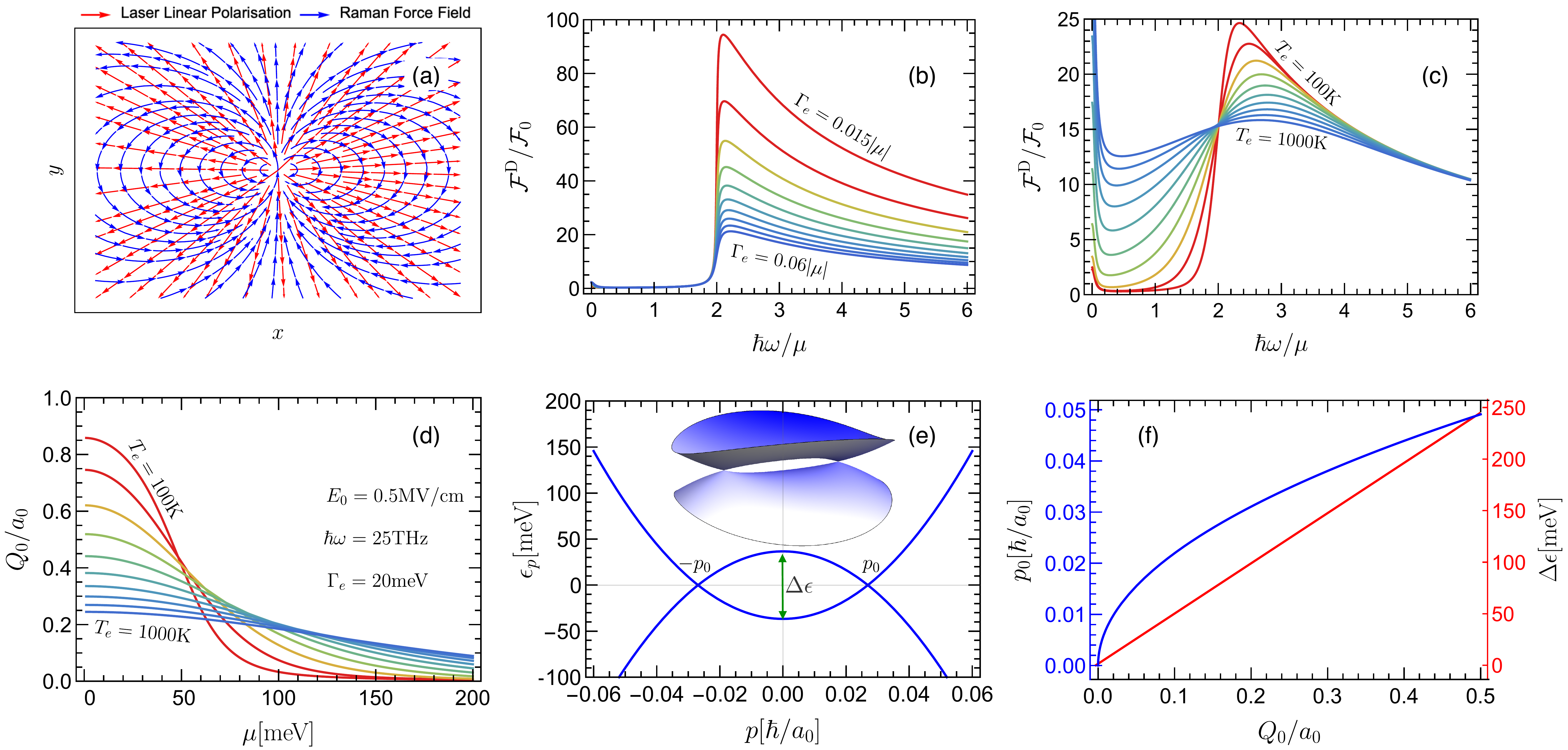}
\caption{{\bf Displacive Raman force and rectified shear displacement in bilayer graphene.} (a) Orientation of the displacive Raman force in comparison to the incident laser polarization . 
 (b) Raman force versus the driving frequency for different values of electronic scattering rate $\Gamma_e$ at zero electronic temperature $T_e=0$. As seen the zero temperature Raman force is finite in the interband regime where $\hbar\omega>2|\mu|$ and for large frequency it scales as ${\cal F}\sim 1/(\Gamma_e \omega)$. (c) Raman force versus the driving frequency for different values of the electronic temperature $T_e$ at $\Gamma_e=0.01 |\mu|$. At finite temperature the displacive force is finite even in the intraband regime $\hbar\omega<2|\mu|$. (d) Rectified shear displacement versus chemical potential for realistic experimental values of frequency $\hbar\omega=25$THz,  electric field strength $E_0=5$MV/cm and electron scattering rate $\Gamma_e=20$meV.
(e) Low-energy dispersion of bilayer graphene under rigid shear displacement $Q_0=0.15a_0$. The original band touching at $p=0$ is lifted, and a pair of Dirac crossing forms at $\pm p_0$. The inset shows the 3D energy dispersion. (f) The band splitting $\Delta\epsilon$ at $p=0$ and the new Dirac crossing location $p_0$ are tuneable by the shear displacement $Q_0$ as depicted in this double-sided plot.}
\label{fig:result}
\vspace{-3mm}
\end{figure*} 

{\em Shear phonon in bilayer graphene.--}
Bilayer graphene consists of two single layers
of graphene sheets offset from each other in the $xy$-plane. The low-energy quasiparticles in BLG follow a two-band Hamiltonian around the corners of the hexagonal Brillouin zone \cite{McCann_prl_2006}
\begin{align}\label{eq:Ham-2band}
\hat H_{\bf p} = -\frac{1}{2m}\{(p^2_x-p^2_y)\hat\sigma_x + 2 \tau p_x p_y \hat \sigma_y \} -\mu\hat I, 
\end{align}
where ${\bf p}=\hbar {\bf k}$ is the momentum vector, $\tau=\pm$ stands for two K and K$'$ valley points, the identity matrix $\hat I$ and Pauli matrices $\hat \sigma_{x,y}$ are in the layer pseudospin basis, and $\mu$ is the chemical potential. The $x$-direction indicates a zigzag orientation of the hexagonal crystal in our convention \cite{Rostami_prb_2013b}. The effective mass is given by $1/2m \approx v^2 /\gamma_1 $ with $v=3\gamma_0 a_0/2\hbar\sim 10^6$m/s. Note that $\gamma_i$s are hopping energies in the lattice model illustrated in Fig.~\ref{fig:schematic}b. We neglect trigonal warping and effective mass asymmetry in the energy dispersion of chiral fermions in BLG. In pristine BLG, two degenerate shear modes correspond to the sliding motion and two Cartesian directions.

In order evaluate Raman force, we  consider the coupling of electrons to one and two photons given by $ \hat j_\alpha = -e \partial_{p_\alpha} \hat H_{\bf p} $ and $\hat \gamma_{\alpha\beta} = - e^2 \partial_{p_\alpha}\partial_{p_\beta} \hat H_{\bf p}$, respectively. In the low-energy model, the couplings of electrons to shear phonons are given by 
\begin{align}\label{eq:electron-phonon}
&(\hat M_x,\hat M_y)\approx M (\tau\hat \sigma_y, \hat\sigma_x),
\nonumber\\  
&(\hat \Theta_{xy}= \hat\Theta_{yx} , \hat\Theta_{yy}=- \hat\Theta_{xx} ) \approx -\Theta (\tau \hat\sigma_x, \hat\sigma_y).
\end{align}
The electron-phonon coupling are obtained using a four-band tight-binding model following the approach given in Refs. \cite{Ishikawa_2006, Cappelluti_prb_2012,Basko_2009} --see Supplemental Material for the detail discussion on electron coupling to shear phonons in BLG. After neglecting electron momentum $p$, we obtain $M=({3 a_0 \gamma_3}/{\sqrt{2} b^2} ) \beta_3$ and $\Theta = ({e\gamma_3}/{\hbar})({3a^2_0}/{ 2\sqrt{2} b^2} ) \beta_3$ where $\beta_3 = - \partial \ln \gamma_3/\partial \ln b$. We set the Gruneisen parameter $\beta_3\sim2$. The vertical hopping derivative ${\partial\gamma_1}/{\partial c}$ does not contribute in the leading order electron-phonon interaction. The photon-electron-phonon coupling is obtained after neglecting electron momentum $p$. 
 
{\em Results and discussion.--}
In analogy to the linear and circular photogalvanic current \cite{Belinicher_1980}, we decompose the displacive Raman force into linear and circular components. The displacive Raman densities are thus formally given by (see Supplemental Material) 
\begin{align}
{\cal F}^{\rm LDR}_a &= \gamma^{\rm LDR}_{abc} (\omega) {\rm Re}[E_b(\omega)E^\ast_c(\omega)],
\nonumber\\
{\cal F}^{\rm CDR}_a &= \gamma^{\rm CDR}_{a} (\omega)  [i{\bf E}(\omega)\times {\bf E}^\ast(\omega)]_z.
\end{align}
For an electric field polarization in the $xy$-plane, the linear displacive Raman (LDR) and circular displacive Raman (CDR) response functions read 
\begin{align}
&\gamma^{\rm LDR}_{abc} (\omega)=- {\rm Re}[\rchi^{\rm R}_{abc}(\omega,-\omega)]/\omega^2,
\nonumber\\
&\gamma^{\rm CDR}_{a} (\omega)=- {\rm Im}[\rchi^{\rm R}_{axy}(\omega,-\omega)]/\omega^2.
\end{align}
For the circular case, we consider a generic elliptical polarization  of the incident laser field ${\bf E}(t) = E_0 \{\cos(\vartheta) \hat{\bf x} \pm i \sin(\vartheta) \hat{\bf y}\} e^{-i\omega t}$.  An elliptical polarization  contains both linear and circular counterparts: $[i{\bf E}(\omega)\times {\bf E}^\ast(\omega)]_z=\pm \sin(2\vartheta)$ and ${\rm Re}[E_b(\omega)E^\ast_c(\omega)] =\delta_{ab} (\delta_{a x}\cos^2\vartheta  +\delta_{a y}\sin^2\vartheta  )$. Considering the inversion and rotational symmetries of the low-energy model, we find non-vanishing tensor elements $ \Lambda_{yyy}=\Lambda_{xxy}=\Lambda_{xyx} = -\Lambda_{yxx} $. Accordingly, the symmetry implies that $\Lambda_{axy}$ is either zero ($a=y$) or real valued ($a=x$) leading to a vanishing circular displacive Raman force in BLG: $\mathbfcal{F}^{\rm CDR}=0$. In order to have the CDR force finite, we need to break rotational symmetry for instance by applying a uniaxial strain. The LDR force contribution owing to an elliptically polarized incident laser reads    
\begin{align}
\mathbfcal{F}^{\rm LDR} = -\hat{\bf y} {\cal F}_0 \Lambda(\bar\omega_1,\bar\omega_2) \cos(2\vartheta). 
\end{align}
Accordingly, the Raman force vanishes for a circularly polarized light $\vartheta=\pm \pi/4$.  
Note that ${\cal F}_0= N_f M (e E_0)^2/(8\pi \mu^2)$ with $N_f=4$ for spin-valley degeneracy. For $\mu=0.2$eV, and $E_0=1$V/nm, we obtain the force density unit ${\cal F}_0\approx 0.8{\rm nN/nm^2}$. 
For linear polarized incident light ${\bf E}(t) = E_0 \{\cos(\theta) \hat{\bf x} + \sin(\theta) \hat{\bf y}\} e^{-i\omega t}$, the Raman force follows 
\begin{align}
&\mathbfcal{F}^{\rm LDR} = {\cal F}_0  \Lambda(\bar\omega_1,\bar\omega_2) 
\{\sin(2\theta) \hat{\bf x} - \cos(2\theta) \hat{\bf y} \}.
\end{align} 
As seen from the above relation, the force vector is not necessarily parallel to the driving electric field. In Fig.~\ref{fig:result}a, the orientation of the Raman force is compared to the incident laser linear polarization . 

The frequency dependence of the Raman force is captured by the dimensionless function $\Lambda(\bar\omega_1,\bar\omega_2) $. There are topologically distinct contributions to the Raman force, which are illustrated diagrammatically in Fig.~\ref{fig:diagrams}. Using equilibrium Green's function method, we analytically calculate Raman response functions at zero electronic temperature $T_e=0$ (see Supplemental Material)
\begin{align}
\Lambda(\bar\omega_1,\bar\omega_2) &= 
\frac{\bar\omega_1+2\bar\omega_2}{(\bar\omega_1+\bar\omega_2)\bar\omega^2_2} \ln\left[\frac{4-\bar\omega^2_1}{4-(\bar\omega_1+\bar\omega_2)^2}\right]
\nonumber\\ &+ 
\frac{\bar\omega_2+2\bar\omega_1}{(\bar\omega_1+\bar\omega_2)\bar\omega^2_1} \ln\left[\frac{4-\bar\omega^2_2}{4-(\bar\omega_1+\bar\omega_2)^2}\right].
\end{align}
Note that $\bar\omega_j= (\hbar\omega_j+i\Gamma_e)/|\mu|$ where $\Gamma_e$ stands for the phenomenological scattering rate of electrons. 
The above expression stems from the triangular diagram and a bubble diagram, as shown in Fig.~\ref{fig:diagrams}a and Fig.~\ref{fig:diagrams}b, respectively. The contribution from the Feynman diagram depicted in Fig.~\ref{fig:diagrams}c vanishes based on our low-energy model analysis. The last diagram shown in Fig.~\ref{fig:diagrams}d, is frequency independent, and its value is fixed by enforcing the gauge invariance where the response to a static homogeneous gauge potential must vanish due to the gauge invariance, i.e. $\rchi^{\rm R}_{abc}(0,0)=0$. 
For the displacive force, we set $\omega_1=-\omega_2=\omega$ and therefore it scales as ${\cal F}\sim 1/\Gamma_e$. In Fig.~\ref{fig:result}b, we illustrate the magnitude of the displacive Raman force versus frequency $\hbar\omega/|\mu|$ for different values of phenomenological scattering rate $\Gamma_e$. As expected, the displacive force is finite in the interband regime when $\hbar\omega>2|\mu|$ and for the large frequency, the force density scales as  ${\cal F}^{\rm D}= {M (eE_0)^2}/({\pi\hbar\omega\Gamma_e})$.  

Since the displacive force and interband optical absorption occur coincidentally, the effect of electronic temperature can not be abandoned. For an intense incident laser, the photo-excited electrons in a metal can reach a quasi-equilibrium state with very high electronic temperature (e.g. $T_e\sim 1000$K) \cite{Andreatta_prb_2019}. 
The results at finite electronic temperature $T_e$ are depicted in Fig.~\ref{fig:result}c,d, which shows a non-zero displacive force even in the intraband regime, and a robust Drude-like tail emerges at low frequency. Although the temperature depends on the optical absorption, we model it as an independent parameter to evaluate the leading-order impact of hot electrons.  

Using ${\bf Q}_{0} \approx \mathbfcal{F}^{\rm D} /(\rho\Omega^2_0)$ and $\rho= c \rho_{gr}$ for the mass density of BLG, with $\rho_{gr}\approx2.267 {\rm kg/cm^3}$ being the three-dimensional graphite density, we estimate the strength of the displacive shear displacement as depicted in Fig.~\ref{fig:result}d. This displacement is robust and tuneable by altering the Fermi energy, incident laser frequency, and laser intensity. 
We find giant values for $Q_0$ in our leading-order theory considering realistic experimental values for laser intensity and carrier doping.  
The saturation value of shear phonon displacement measured to be around $Q_{\rm max}\sim 8$pm in layered WTe$_2$ using intense infrared laser with electric field strength $E_0\sim7.5$MV/cm\cite{Sie_nature_2019} and $E_0\sim10$MV/cm\cite{Ji_acsnano_2021}.  Although the earlier experiment \cite{Sie_nature_2019} does not support a Raman mechanism, the later \cite{Ji_acsnano_2021} measures a linear power dependence of the shear displacement consistent with the Raman force $ {\cal F}^{\rm D} \propto E^2_0$. Such linear power dependence is also reported in Ref. \cite{Zhang_prx_2019}. For a realistic doping $\mu=200$meV, electronic temperature $T_e=1000$K and scattering rate $\Gamma_e= 20$meV in bilayer graphene, we find $Q_0\sim 0.13a_0\sim 12$pm for $E_0\sim5$MV/cm and $\hbar\omega =25$THz that can grow up to $Q_0\sim 50$pm for $E_0\sim 10$MV/cm. This value is clearly immense, and it suggests the need for higher-order corrections in the electric field and phonon anharmonicity leading the saturation of displacement.  

The rigid shear displacement (frozen shear phonon) induces a perturbation to the electronic Hamiltonian according to Eq.~(\ref{eq:Ham}) that for instance at the K-point ($\tau=+$) it reads $ \delta \hat H = M (Q_{0,x}\hat \sigma_y+Q_{0,y}\hat\sigma_x)$. 
As a result of this frozen shear phonon, the band touching point at $p=0$ splits by $\Delta\epsilon=2MQ_0$ and a Dirac crossing pair forms at ${\bf p}_0 =\pm p_0(\cos\phi_0,\sin\phi_0)$ in the Cartesian coordinates where we find (see Fig.~\ref{fig:result}e) 
\begin{align}
p_0 = \sqrt{2m M Q_0}~~~,~~~\tan(2\phi_0) = Q_{0,x}/Q_{0,y}. 
\end{align}
For linear polarized incident laser, the rectified shear displacement is ${\bf Q}_0 = Q_0( \sin(2\theta), -\cos(2\theta))$ with $\theta$ being the incident laser polarization  angle. Therefore, we find $\phi_0 = -\theta$ which implies that the new Dirac crossing points are aligned to the incident laser polarization .
The energy splitting $\Delta \epsilon$ and the position of the crossing point $p_0$ depend on the rigid shear displacement as illustrated in Fig.~\ref{fig:result}f. In principle, this change in the dispersion is large enough to be experimentally measured utilising the angle-resolved photoemission spectroscopy (ARPES). 

The proposed mechanism of displacive Raman force can strongly impact the Raman-active shear phonon dynamics in twisted systems, particularly in trilayer graphene. The theory can be generalized to investigate displacive coherent dynamics of other collective modes, such as magnons \cite{Neto_prb_2005} and superconducting Higgs mode \cite{Cea_prb_2019}, driven by a rectified light-induced force field. In future studies, we will develop a higher-order Raman force mechanism to manipulate chiral valley phonons \cite{Lifa_prl_2015,rostami2022strain} in hexagonal 2D materials. In a separate study, we will discuss the saturation of rigid shear displacement using a higher-order  Raman force scheme.

%{\em Acknowledgment.--}  
This work was supported by Nordita and the Swedish Research Council (VR 2018-04252). 
Nordita is partially supported by Nordforsk. I am grateful to E. Cappelluti, J. Weissenrieder, and F. Guinea for constructive discussion and helpful feedback.

\bibliography{refs.bib}  

\end{document}